\documentstyle[aps]{revtex}
\newcommand{\be}{\begin{equation}} \newcommand{\ee}{\end{equation}}
\newcommand{\bey}{\begin{eqnarray}} \newcommand{\eey}{\end{eqnarray}} 

\title{On the determination of CP violating Majorana phases} 
\author{M K Samal\footnote{e-mail:mks@iiap.ernet.in}} 
\address{Institute of Physics, Sachivalaya Marg, Bhubaneswar 751 005,
India.}

\begin{document} 
\maketitle 

\begin{abstract}
The determination of CP violating phases in the Majorana neutrino mixing matrix
using phenomenological constraints from neutrino oscillations, neutrinoless
double beta decay, ($\mu^- , e^+$) conversion and few other processes is
discussed. We give the expressions for the phases in terms of the mixing angles
and masses consistent with the recent data from Kamiokande. 
\end{abstract}

\vskip 3mm 

At present, experiments involving solar neutrinos and/or atmospheric neutrinos and
the necessity of a hot component in the favoured mixed dark matter scenario indicate
that neutrinos possess a small mass.  The see-saw mechanism\cite{ramond}
attributes the smallness of neutrino mass (compared to that of charged leptons) to
its Majorana nature, hence to its electric charge neutrality.  Majorana neutrinos
satisfy the Majorana self-conjugate conditions: 
\be
\nu^C_j \equiv \nu_j; \:\:\: (j=1,2,3), \label{majcond}
\ee
which do not permit any arbitrary rotation of the neutrino fields. Consequently,
in contrast to the quark case, the analogous unitary neutrino mixing matrix defined
by
\begin{equation}
\nu_\alpha = \sum_j U_{\alpha j} \nu_j; \:\:\: (\alpha =e, \mu, \tau),
\end{equation}
contains three CP-violating phases. Then the mixing matrix can be parametrized
as\cite{bilenky}:
\be
\left(
\begin{array}{ccc} 
c_{1}& s_{1} c_{3}e^{i \alpha}&s_{1} s_3 e^{i \beta}\\
-s_{1}c_{2} e^{-i \alpha}&c_{1}c_{2}c_3 - s_{2} 
s_{3} e^{i \delta}  &(c_{1} c_{2} s_3 + s_2 c_3 e^{i \delta}) e^{- i
(\alpha-\beta)}\\ 
-s_{1}s_{2} e^{-i \beta}& (c_{1} s_{2} c_3 + c_2 s_3 e^{i \delta}) e^{ i
(\alpha-\beta)}&c_{1}s_{2}c_3 - c_{2}  c_{3} e^{i \delta}
\end{array}
\right)
\ee
where $s_{j}\equiv \sin{\theta_{j}}$, $c_{j}\equiv \cos{\theta_{j}}$ and $\theta_j$
are the neutrino mixing angles. The phase $\delta$ is the usual CP violating phase
one encounters if neutrinos are considered to be Dirac particles like quarks, but
the extra two CP-violating phases $\alpha$ and $\beta$ are intrinsic to Majorana
nature of neutrinos. In a different phase convention for the Majorana
condition\cite{malaapi}
\be
\nu^C_1 \equiv \nu_1; \:\: \nu^C_2 \equiv e^{2 i \alpha} \nu_2; \:\: \nu^C_3  \equiv
e^{2 i \beta} \nu_3, \label{majcond2}
\ee
it was shown that all the CP violation characteristic of Majorana nature can be put
into the Majorana condition (\ref{majcond2}) such that the matrix $U$ has only CP
violating KM phase $\delta$. But in such a case some elements of $U$ could be purely
imaginary even if CP is conserved. In any case, physical quantities should be
independent of the parametrization.  In this note we discuss about the possible
determination of these phases (we use phase convention (\ref{majcond})) from the
phenomenological constraints available in neutrino sector. 

CP-violating effects in the KM model with Majorana neutrinos can be catagorised as
follows:

a) CP-violation phenomena are independent of CP-violating Majorana phases. 

This is identical to the case of Dirac particles and not only the charged lepton
fields but also the neutrino fields are defined upto arbitrary phases. Any measure
of CP-violation that is independent of the rephasing ambiguity of leptonic fields
can be shown to be proportional to the imaginary parts of the rephasing
invariant\cite{jarlskog}
\be
J_{\gamma k}= U_{\alpha i}U_{\beta j}U^*_{\alpha j}U^*_{\beta i};\:
(\alpha,\beta,\gamma \:\; {\rm and}\:\; i,j,k \:\; {\rm cyclic}), 
\ee
or of the products of $J_{\gamma k}$. For three generations, there are nine of such
invariants $J_{\gamma k}$ but the unitarity of $U$ makes all of them have the same
imaginary component including the sign
\be
Im \; [J_{\gamma k}]= J =c_1 c_2 c^2_3 s_1 s_2 s_3 \sin{\delta}.
\ee

The phase $\delta$ can be determined once one knows $J$ and the mixing angles.
Recently there have been many studies\cite{cpstud} where extracting $J$ from the
long-baseline three-flavour neutrino oscillations experiments is discussed. 
Existing neutrino anomalies imply\cite{smirnov} three different scales of
neutrino mass squared differences $\delta m^2 \equiv (m^2_j-m^2_k)$:
\bey
\delta m^2_{LSND}& \sim & (0.2 - 2)\:\: {\rm eV}^2,\nonumber\\
\delta m^2_{ANA}& \sim & (0.3 - 3) \times 10^{-2}\:\: {\rm eV}^2,\nonumber\\
\delta m^2_{SNP}& \sim & (0.3 - 1.2) \times 10^{-5}\:\: {\rm eV}^2,\label{nmdata}
\eey
which satisfy
\be
\delta m^2_{LSND}  >> \delta m^2_{ANA} >>\delta m^2_{SNP}  ,
\ee
where, $\delta m^2_{LSND}$, $\delta m^2_{ANA}$, $\delta m^2_{SNP}$ correspond to
the mass squared differences involved in the LSND measurement, atmospheric neutrino
anomaly and solar neutrino puzzle respectively. The mass scale that provides the
desired $20 \%$ HDM component of the mixed dark matter of the universe,
\be
\delta m^2_{HDM} \sim (1-50) \:\:  {\rm eV}^2,
\ee
can accomodate the LSND range. But it is impossible to reconcile all the anomalies
with only three neutrinos without extra assumptions because the data (\ref{nmdata}) 
is not commensurate with the obvious relation $ \delta m^2_{12} + \delta m^2_{23}
=\delta m^2_{31}$. In recent studies, these anomalies have been reconciled by
assuming that $\delta m^2_{LSND}= \delta m^2_{ANA}$ (or $\delta m^2_{SNP}= \delta
m^2_{ANA}$) or by ignoring at least one anomaly or by introducing additional sterile
neutrinos. We choose the highest neutrino mass scale to be $O(1\: {\rm eV})$, which
is appropriate for cosmological hot dark matter and the lower mass
scale to be $\delta m^2 \simeq 10^{-2} \:\; {\rm eV}^2$ as we are interested in
studies of CP violation in long baseline experiments which are motivated by the
atmospheric neutrino mass scale.
Then the solar neutrino puzzle is solved through the mixing of a sterile neutrino of
mass $m_s \sim (2-3) \times 10^{-3} \: $ eV with $\nu_e$ via MSW conversion. The
strong nucleosynthesis bound on the number of light neutrino species is satisfied
since the sterile neutrino has the parameters of the solar neutrino puzzle.
The masses of $\mu_\mu$ and $\nu_\tau$ are in the range of $2-3$ eV and they provide
the hot component of the mixed dark matter. The atmospheric neutrino anomaly is
understood  through $\nu_\mu \rightarrow \mu_\tau$ oscillations with maximal mixing
between $\nu_\mu$ and $\nu_\tau$ that form a pseudo-Dirac neutrino. At the same
time, $\overline{\nu_\mu} \rightarrow \overline{\nu_e}$ mixing can explain the LSND
result. In this scenario, the neutrino masses satisfy $m_1 < m_s << m_2 \sim m_3$.

With the above mass hierarchy chosen, the oscillation probability in vacuum for the
long-baseline experiments can be written as\cite{mina}
\be
P(\nu_\beta \rightarrow \nu_\alpha)=A_{\beta \alpha} +B_{\beta \alpha} (1-
\cos{\phi_{jk}}) +C_{\beta \alpha} \sin{\phi_{jk}},
\ee
with $\phi_{jk}= \delta m^2_{jk} L/ 2E$, where $L$ and $E$ denotes the path-length
of the baseline and the energy of the neutrino respectively. The coefficients
$A_{\beta \alpha}$, $B_{\beta \alpha}$ and $C_{\beta \alpha}$ are constants which
depend on the mixing angles and the CP violating phase $\delta$. In particular the
coefficient $C_{\beta \alpha}$ is $2 J$.
>From the structure of the oscillation probability it is easy to observe that
\be
P_{\nu_\ell} \left(x_\mu, x_\nu; \phi_{jk}= \frac{3
\pi}{2}\right)-P_{\nu_\ell} \left(x_\mu, x_\nu; 
\phi_{jk}= \frac{\pi}{2}\right)=4 J.  
\ee
Therefore measurement of $P_{\nu_\ell} \left(x_\mu, x_\nu; \phi_{jk}= \frac{3
\pi}{2}\right)$ and $P_{\nu_\ell} \left(x_\mu, x_\nu; \phi_{jk}=
\frac{\pi}{2}\right)  $ can yield $J$. Since $\phi_{jk}$ are functions of $L$ and
$E$ for a chosen $\delta m^2_{jk}$, the measurement of the above difference in
probabilities can be done either varying $L$ or $E$, or both\cite{mina2}. 
Another way of estimating $J$ is to note
that a direct measure of CP violation in three-neutrino oscillations is the unique
difference of the transition probabilities between CP-conjugate
channels\cite{sandip}: 
\bey
\Delta P & \equiv & P(\overline{\nu}_\mu - \overline{\nu}_e) - 
P(\nu_\mu - \nu_e) = P(\nu_\mu - \nu_\tau) - P(\overline{\nu}_\mu - 
\overline{\nu}_\tau) = P(\overline{\nu}_e - \overline{\nu}_\tau)  - 
P(\nu_e - \nu_\tau) \nonumber\\
&= & 4 J S,
\eey 
where the contribution from the mixing matrix is in terms of $J$ and the squared
mass differences contribute through the term $S=\sum_{j < k} \sin \phi_{jk}$. Hence
one can find $J$ from the long baseline neutrino oscillation
experiments with the proper choice of the mass hierarchy for neutrinos. 

Once the neutrino mass scales are fixed, one can discuss the pattern of the $3
\times 3$ neutrino mixing matrix.  Using the Kamiokande binned data for
sub-GeV and multi-GeV atmospheric neutrinos, Yasuda has found\cite{yasuda} the
optimum set of three flavour neutrino oscillation parameters within the constraints
of the reactor experiments as:
\bey
(\delta m^2_{21}, \delta m^2_{32}, \theta_{1}, \theta_{3}, \theta_{2})&=&( 7.4
\times 10^{-1} \: {\rm eV}^2, 2.6 \times 10^{-2} \:{\rm eV}^2, 2^{\circ}, 3^{\circ},
45^{\circ})\nonumber\\ &&{\rm and}\:  ( 2.6 \times 10^{-2} \:{\rm eV}^2, 7.4 \times
10^{-1} \: {\rm eV}^2, 0^{\circ}, 87^{\circ}, 46^{\circ}) 
\eey
We will use this data for our analysis because we are concerned with the long
baseline experiments that have been motivated by the atmospheric neutrino data.
Thus once $J$ is extracted from the experiments and the pattern of mixing angles is
chosen, one can estimate the phase $\delta$.

b) CP-violation effects depend on the Majorana phases $\alpha$ and $\beta$. 

The only phase freedom in this case lies in the charged lepton fields and the
rephasing invariant combination\cite{pal} of the mixing matrix that contains
Majorana phases and has the imaginary part turns out to be of the form
\be
K_{\gamma k}= m_i m_j U_{\alpha i}U^*_{\beta j}U^*_{\alpha j}U_{\beta i};\:
(\alpha,\beta,\gamma \:\; {\rm and}\:\; i,j,k \:\; {\rm cyclic}), 
\ee
or of combinations of $K_{\gamma k}$. The imaginary parts of $K_{\gamma k}$ vary
from case to case.
Where does one look for these phases ? Let us consider the following processes:

i) The neutrino flavour oscillations:

It was shown\cite{bilenky2} that these phases do not appear in the neutrino
flavour oscillation probabilities. Hence no neutrino oscillation experiments can
distinguish between Dirac and Majorana neutrinos.

ii) The neutrino-antineutrino oscillations: 

In contrast to the quark sector, a CP-violating phase remains in $U$ even in the
two-generations case for Majorana neutrinos. In this case the CP-violating Majorana
phase enters\cite{antinue} directly into the argument of neutrino-antineutrino
oscillation probability and thus in principle a measurable quantity. But in practice
its effect is hard to measure since it shows up in lepton-number violating processes
suppressed by a factor $m_\nu/E_\nu$ compared to the usual weak interaction
processes.

iii)The $ \mu \rightarrow e \: \nu_1 \: \nu_2$ process: 

Due to the Majorana nature of neutrinos there is an additional Feynman diagram
in this case whose interference gives rise to a term\cite{doietl} proportional to
\be
m_i m_j U_{e 1}U^*_{\mu 2}U^*_{e 2}U_{\mu 1}= -m_1 m_2 s_1^2 c_1 c_2 c_3 ( c_1 
c_2 c_3 -s_2 s_3 e^{ i \delta})e^{-2 i \alpha}. \label{clean}
\ee
However the CP-violating effect associated with the imaginary part of
eqn.(\ref{clean}) is suppressed by a factor of $(m_1 m_2/m^2_\mu)$ and thus is
very small. Otherwise this process could have been one of the cleanest ways to check
for the contribution of $\alpha$ to the CP-violating effect through the
presence of $\sin{2 \alpha}$ in eqn.(\ref{clean}) since the mixing angles
satisfy $c_1 c_2 c_3 >> s_2 s_3$ for one allowed range of parameters. 

iv) The $ \mu \rightarrow 3 e$ process: 

In this case, the CP asymmetry resulting from the different interference effects
between particle and anti-particle channels is proportional to
a rephasing invariant function of mixing matrix elements as:
\bey
\Delta & =& \frac{\Gamma(\mu \rightarrow 3 e)-\Gamma(\overline{\mu}
\rightarrow 3 \overline{e})}{\Gamma(\mu \rightarrow 3 e)+\Gamma(\overline{\mu}
\rightarrow 3 \overline{e})} \nonumber\\
& & \propto  Im \left( [\sum_{j} m_j U^2_{ej}][\sum_{j} m_j
U^*_{ej} U_{\mu j}][\sum_{k} m_k U^*_{ek} U_{\mu k}] \right),
\eey
that can be used to constrain the phases $\alpha$ and $\beta$. But it was
shown\cite{cheng} that CP-violating phenomena through loop effects, which arise from
charged gauge currents are unobservable due to GIM suppression. However it was
pointed out that one could hope to probe CP-violating effects in $\mu \rightarrow 3
e$ as they can proceed through the charged Higgs boson exchange. This difference
between the two cases arises because in the Higgs boson exchange, the VEV of the
charged Higgs field i.e.  $v$ replaces $m_W$ in the GIM suppression factor
$m_\nu/m_W$. This was a tremendous improvement with the then existing
bound\cite{dear} $v < 9 $ keV. But in the light of LEP data this bound has changed
to $v > 43.5$ GeV and consequently, the Higgs boson exchange process is as much
suppressed as the charged gauge boson exchange process. Thus this process is an
unlike candidate to study these CP-violating phases.

v) The neutrinoless double beta decay ($(\beta \beta)_{0 \nu}$):

The neutrinoless double beta decay of a nucleus $(A, Z)$,
\be
(A, Z) \rightarrow (A, Z+2) + 2 e^-
\ee
violates lepton number by two units and is possible only if neutrinos are Majorana
particles. The decay rate is, in the absence of right-handed currents, proportional
to the effective Majorana mass $<m_\nu>^2$ which is given, in the case of light
neutrinos ($m_\nu < 1$ MeV) by
\be
<m_\nu>= | \sum_{j} m_j U^2_{ej}|,
\ee
where $m_i$ is the mass corresponding to the mass eigenstate of the i-th generation
Majorana neutrino. In the special case of two generations of Majorana neutrinos, 
\be
<m_\nu>= | m_1 c_1^2 + m_2 s_1^2 e^{i 2 \alpha}|,
\ee
where $\alpha$ is the Majorana phase, on which a bound was obtained\cite{kim}
independent of lepton mixing angle as
\be
\sin^2{\alpha} \le 0.078; \:\: |\alpha| < 16^{\circ},
\ee
using the experimental input $<m_\nu>\: < 5.6 \:\;{\rm eV}\:\: {\rm and}\:\: m_1 <
20 \:\; {\rm eV}$.  Here the neutrino mass $m_1$ was considered to be approximately
equal to $m(\nu_e)$ by assuming $m_2 >> m_1$.
But in the light of recent experimental data\cite{pdg} on these quantities, 
$<m_\nu> < 0.56 \:\;{\rm eV}\:\: {\rm and}\:\: m_1 < 15 \:\; {\rm eV}$
the above bound is found to change as:
\be
\sin^2{\alpha} \le 0.001; \:\: |\alpha| <2^{\circ} .
\ee

If there is a MeV Majorana neutrino apart from one light neutrino, the effective
Majorana mass $<m_\nu>$ takes a value\cite{halprin}
\be
<m_\nu>= | m_1 c_1^2 + F(m_2, A) m_2 s_1^2 e^{i 2 \alpha}|,
\ee
with the $A$ dependance through the function
\be
F(m_2, A)= <1/r>^{-1} \: <\exp(-m_2 r)/r>,
\ee
where $r$ corresponds to the distance of the two nucleons in the nucleus undergoing
$(\beta \beta)_{0 \nu}$ decay and the average is with respect to the two nucleon
correlation function appropriate for the nucleus.
In this case we obtain
\be
\sin^2{\alpha}= \frac{(m_1 c^2_1 + m_2 s^2_1 F (m_2, A))^2- <m_\nu>}{4 
m_1 m_2 c^2_1 s^2_1  F (m_2, A)},
\ee 
which may be investigated by comparing half-life limits of different isotopes.

For the three generations of light neutrinos, the effective Majorana mass is given
by
\be
<m_\nu>= | m_1 c_1^2 + m_2  c_3^2 s_1^2 e^{i 2 \alpha} +
m_3 s_3^2 s_1^2 e^{i 2 \beta}|,
\ee
where neither the mixing angle $\theta_2$ nor the KM phase $\delta$ enters the
expression and thus allows to determine $\alpha$ and $\beta$ without interference. 
In an analysis\cite{nishi} of CP-violation effects for the three-generations case,
with use of the assumption $<m_\nu>\: <m_1$, the bound on the phases $\alpha$ and
$\beta$ were given as
\be
f_- \le \sin^2{\alpha '} \le f_+,\:\:\:\: g_- \le \sin^2{\beta '} \le g_+,
\ee
where $\alpha ' = \pi /2-\alpha, \beta '=\pi/2-\beta $ and the upper and lower
limits $f_\pm$ and $g_\pm$ are given in terms of mixing
angles, neutrino masses and $<m_\nu>$ by
\bey
f_\pm&=& \frac{(<m_\nu> \pm m_3 s_3^2 s_1^2 )^2 -(m_1 c_1^2 - m_2 c_3^2 s_1^2 )^2}{4
m_1 m_2 c_1^2 c_3^2 s^2_1}, \nonumber\\
g_\pm&=& \frac{(<m_\nu> \pm m_2 c_3^2 s_1^2 )^2 -(m_1 c_1^2 - m_3 s_3^2 s_1^2 )^2}{4
m_1 m_3 c_1^2 s_3^2 s_1^2}.
\eey
Since the cases when $ \sin^2{\alpha '}$ and $ \sin^2{\beta '}$ take values of 0 or
1 both correspond to
CP-conserving case\cite{kim,doikw} the above bounds are further restricted to
\be
0 < f_- \le \sin^2{\alpha '} \le f_+ <1,\:\:\:\: 0<g_- \le \sin^2{\beta '} \le
g_+<1.
\ee
In a recent study\cite{nishi2} these bounds have been used to constrain the neutrino
mixing angles from the observed data of $<\mu_\nu>$. 
But we use the data from neutrino masses and mixing to determine the phases as
follows:

Case I: $( \theta_{1}= 2^{\circ}, \theta_{3}=3^{\circ}, \theta_{2}=45^{\circ})$

Using the approximation $c_1 \simeq 1$, $s_1 <<1$, $c_3 \simeq 1$, $s_3 <<1$, $c_2
\sim s_2$ the expression for the weighted neutrino mass in this case can be written
as:
\be
<m_\nu>= | m_1 c_1^2 + m_2  c_3^2 s_1^2 e^{i 2 \alpha}|,
\ee 
from which we obtain
\be
\sin^2{\alpha}= \frac{(m_1 c^2_1 + m_2 s^2_1 c^2_3)^2- <m_\nu>}{4 
m_1 m_2 c^2_1 s^2_1 c^2_3}.
\ee 

Case II: $( \theta_{1}= 0^{\circ}, \theta_{3}=87^{\circ}, \theta_{2}=46^{\circ})$

Using the approximation $c_1 =1$, $s_1=0$, $s_3 \simeq 1$, $c_3 <<1$, $c_2
\sim s_2$ the expression for the weighted neutrino mass in this case can be written
as:
\be
<m_\nu>= m_1 c_1^2,
\ee
and nothing can be concluded about the phases.

vi) The ($\mu^- , e^+$) conversion:

The ($\mu^- \: e^+$) conversion process\cite{doi3} is very similar to the $(\beta
\beta)_{0 \nu}$ process and the conversion rate is proportional to the `weighted'
neutrino mass given by
\be
<m_\nu>_{\mu e}= | \sum_{j} m_j U^2_{ej} U^2_{\mu j}|= | m_1 c_1 c_2 - m_2 c_3 e^{i
2 \alpha}(c_{1}c_{2}c_3 - s_{2} s_{3} e^{i \delta}) - m_3 s_3 e^{i 2 \beta}(c_{1}
c_{2} s_3 + s_2 c_3 e^{i \delta})| s^2_1.
\ee

Note that unlike the case of neutrinoless double beta decay, all the three phases
appear in the expression. Hence if one puts the value of KM phase $\delta$ estimated
from the long-baseline neutrino oscillation experiments then it is possible to give
the bounds on the phases $\alpha$ and $\beta$ as it was done in the case of $(\beta
\beta)_{0 \nu}$.
But we will use the approximations based on the pattern of mixing angles to estimate
the CP-violating phases as follows:

Case I: $( \theta_{1}= 2^{\circ}, \theta_{3}=3^{\circ}, \theta_{2}=45^{\circ})$

Using the approximation $c_1 \simeq 1$, $s_1 <<1$, $c_3 \simeq 1$, $s_3 <<1$, $c_2
\sim s_2$ the expression for the weighted neutrino mass in this case can be written
as:
\be
<m_\nu>_{\mu e}= s^2_1 c_1 c_2 | m_1 - m_2 c^2_3 e^{i 2 \alpha}|,
\ee
from which it is easy to obtain
\be
\sin^2{\alpha}= \frac{<m_\nu>_{\mu e}-c^2_1 c^2_2(m_1-m_3 c^2_3)^2
}{4 m_1 m_2 c^2_1 c^2_2 c^2_3}.
\ee 

Case II: $( \theta_{1}= 0^{\circ}, \theta_{3}=87^{\circ}, \theta_{2}=46^{\circ})$

Using the approximation $c_1 =1$, $s_1=0$, $s_3 \simeq 1$, $c_3 <<1$, $c_2
\sim s_2$ the expression for the weighted neutrino mass in this case can be written
as:
\be
<m_\nu>_{\mu e}= s^2_1 c_1 c_2 | m_1 - m_3 s^2_3 e^{i 2 \beta}|,
\ee
which yields
\be
\sin^2{\beta}= \frac{<m_\nu>_{\mu e}-c^2_1 c^2_2(m_1-m_3 s^2_3)^2
}{4 m_1 m_2 c^2_1 c^2_2 s^2_3}.
\ee 

To conclude, we have discussed about the possible determination of the CP-violating
Majorana phases using phenomenological constraints from the pattern of neutrino
masses and mixing, neutrinoless double beta decay, the $(\mu^- , e^+)$ conversion
process and few other lepton number violating processes. We have also given the
expression for the phases in terms of the mixing angles and masses consistent with
the recent data from Kamiokande. 

\section*{Acknowledgements}

I am thankful to the High Energy Physics Group at ICTP for providing local
hospitality and necessary support for a visit during which this work was done. I
thank Prof.  A. Yu. Smirnov for discussions.

\end{document}